\documentclass[11pt]{article}
\usepackage{amsmath,amssymb,color,graphics,epsfig,cite}

\usepackage{amsfonts}

\newcommand{\hoch}[1]{$\, ^{#1}$}

\newcommand{\be}{\begin{equation}}
\newcommand{\ee}{\end{equation}}
\newcommand{\bea}{\setlength\arraycolsep{2pt} \begin{eqnarray}}
\newcommand{\eea}{\end{eqnarray}}
\newcommand{\nn}{\nonumber}

\def\ft#1#2{{\textstyle{\frac{\scriptstyle #1}{\scriptstyle #2} } }}
\def\fft#1#2{{\frac{#1}{#2}}}

\def\0{{\sst{(0)}}}
\def\1{{\sst{(1)}}}
\def\2{{\sst{(2)}}}
\def\3{{\sst{(3)}}}
\def\4{{\sst{(4)}}}
\def\5{{\sst{(5)}}}
\def\6{{\sst{(6)}}}
\def\7{{\sst{(7)}}}
\def\8{{\sst{(8)}}}
\def\sst#1{{\scriptscriptstyle #1}}
\def\oneone{\rlap 1\mkern4mu{\rm l}}

\begin{document}

\begin{center}
{\Large {\bf Consistent Warped $\mathbb R\times T^{1,1}$ Reduction of Heterotic Supergravity
}}

\vspace{20pt}

{\large Liang Ma\hoch{1} and H. L\"u\hoch{1,2}}

\vspace{10pt}

{\it \hoch{1}Center for Joint Quantum Studies and Department of Physics,\\
School of Science, Tianjin University, Tianjin 300350, China }

\bigskip

{\it \hoch{2}The International Joint Institute of Tianjin University, Fuzhou,\\ Tianjin University, Tianjin 300350, China}

\vspace{40pt}

\underline{ABSTRACT}
\end{center}

We find a curious duality between the bosonic sector of heterotic supergravity and the noncritical bosonic string with conformal anomaly. This allows us to map the Mink$_4\times S^3\times S^3$ vacuum of the bosonic string to a new warped $\mathbb R\times (T^{1,1}\times$Mink$_4)$ vacuum of the heterotic theory, which turns out to be supersymmetric, preserving $1/4$ of the supersymmetry. We perform consistent dimensional reduction of the heterotic theory on the warped $\mathbb R\times T^{1,1}$ internal space and obtain ${\cal N}=1$, $D=4$ supergravity, with one complex scalar multiplet and $SU(2)\times SU(2)$ Yang-Mills multiplets, where the gauge symmetry is originated from the isometry group of $T^{1,1}$.

\vfill{\footnotesize maliang0@tju.edu.cn\ \ \ mrhonglu@gmail.com}

\thispagestyle{empty}
\pagebreak

\tableofcontents
\addtocontents{toc}{\protect\setcounter{tocdepth}{2}}

\newpage

\section{Introduction}

Interpreting gauge symmetries as having a geometric origin in higher dimensions was pioneered by Kaluza and Klein \cite{Kaluza:1921tu,klein}. Pauli later envisioned the possibility of obtaining a four-dimensional theory that encodes the information of the isometry group $SO(3)$ of the internal two-sphere in a six-dimensional Mink$_4\times S^2$ spacetime \cite{ORaifeartaigh:1998ddh,Straumann:2000zc}. His attempt failed since Kaluza-Klein dimensional reduction of gravity on spheres is far subtler than on a circle. The reductions are generally inconsistent \cite{Duff:1983gq,Duff:1984hn,Duff:1985jd}. However, arguments \cite{Duff:1983gq,Duff:1984hn,Duff:1985jd,deWit:1984nz} for consistency in $S^7$ reduction of 11-dimensional supergravity \cite{Cremmer:1978km} were proposed.  The corresponding consistent $S^7$ reduction ansatz was given in \cite{deWit:1986oxb}, although the complete and explicit ansatz for the reduction of the 3-form field was obtained later in \cite{deWit:2013ija,Godazgar:2013dma,Godazgar:2013pfa,Godazgar:2015qia}. Both before and after being motivated by the AdS/CFT correspondence \cite{Maldacena:1997re}, many explicit examples of consistent sphere reductions from $D=11$ and $D=10$, with complete and partial reduction Ans\"atze, were obtained \cite{Pilch:1984xy,Gunaydin:1984fk,Kim:1985ez,Cvetic:1999xp, Nastase:1999cb, Lu:1999bc,Cvetic:1999un,Lu:1999bw,Cvetic:1999au,Nastase:1999kf, Cvetic:2000nc,Azizi:2016noi}. Group theoretical arguments based on the enhanced global symmetries of supergravities in the Kaluza-Klein torus reductions were given to test when it might be consistent to perform sphere reductions \cite{Cvetic:2000dm}. In particular, the arguments suggest that it is consistent to perform $S^3$ or $S^{D-3}$ reductions of the low energy effective theory of the bosonic strings in general $D$ dimensions, and indeed the explicit Ans\"atze were given \cite{Cvetic:2000dm}.

However, gauged supergravities in lower dimensions arising from consistent sphere reductions typically have a negative cosmological constant that is of the same order as the Yang-Mills gauge coupling. Whilst this may be a desirable feature in the AdS/CFT correspondence, it makes these theories less desirable in the areas of phenomenology in both particle physics and cosmology. A notable counterexample is the ${\cal N}=(1,0)$ chiral Salam-Sezgin model in six dimensions, which admits the Mink$_4\times S^2$ vacuum \cite{Salam:1984cj}. It turns out that there exists a consistent $S^2$ reduction that gives rise to ${\cal N}=1$, $D=4$ supergravity coupled to a complex scalar and $SU(2)$ Yang-Mills multiplets, but without a cosmological constant \cite{Gibbons:2003gp}. This provides the first concrete solution to Pauli's vision that the $SU(2)$ Yang-Mills gauge symmetry in asymptotically flat spacetime has a geometric origin in the isometry group of $S^2$.  These features provide applications of supergravity through sphere reductions in cosmology \cite{Aghababaie:2003ar}.

The bosonic Lagrangian of the Salam-Sezgin model can also be obtained from the generalized warped Kaluza-Klein reduction of $D=7$ supergravity \cite{Kerimo:2003am}; however, the resulting theory is not ${\cal N}=(1,0)$ chiral supergravity, but rather a variant of ${\cal N}=(1,1)$ supergravity \cite{Kerimo:2003am,Kerimo:2004md}. More supergravities with Mink$\times$sphere were constructed using the analogous technique \cite{Kerimo:2004qx}. These theories provide further possibilities for obtaining more supergravities, where Yang-Mills fields have geometric origins as in the case of gauged supergravities, but now with Minkowski vacua.

There are two types of consistent dimensional reductions. One type is the Pauli type, where one gets all the Yang-Mills fields associated with the full isometry group of the internal space. The aforementioned $S^7$ reduction of 11-dimensional supergravity is one such example, and its consistency is not guaranteed in general and requires miraculous properties, which, in the context of supergravities or gauged supergravities, can be understood in the framework of generalized and exceptional geometry \cite{Lee:2014mla,Hohm:2014qga, Baguet:2015iou,Cassani:2019vcl}. (The formalism can be extended \cite{Hassler:2020rlq} to include theories with pseudo-supersymmetry \cite{Lu:2011nga,Lu:2011zx,Lu:2011ku,Lu:2011vk}.) The Kaluza-Klein circle reduction is another type, which is a special case of the more general DeWitt type \cite{DeWitt}. If the internal space has an isometry group or one of its subgroups $G$, the reduction is guaranteed to be consistent if the reduction ansatz contains all the singlets of $G$. In a group manifold of group $G$, the metric has $G_L\times G_R$ isometry. It is thus always possible to perform DeWitt type of reduction, keeping all the singlets of $G_L$ subgroup, (see, e.g.~\cite{Lu:2006ah}.) This singles out Kaluza-Klein on $S^1$ or its product $(S^1)^n=T^n$. In addition, $S^3$ is special, since it is not only a coset like general $S^n$, but also a group manifold of $SU(2)$, with the isometry group $SO(4)\sim SU(2)_L\times SU(2)_R$. It is thus consistent to reduce any gravity theory on $S^3$ or its product spaces, as long as the reduction ansatz involves all the singlets constructed from the left-invariant $SU(2)$ 1-forms. The reduction ansatz is as straightforward as the Kaluza-Klein torus reduction.

One application of the DeWitt reduction is that it is consistent to perform the nontrivial Pauli reduction of a Kaluza-Klein theory on $S^2$ \cite{Cvetic:2000dm}. This is because $S^3$ can be viewed as a $U(1)$ bundle over $S^2$ and the DeWitt reduction ansatz preserves this structure. This feature can explain the consistency of the Pauli $S^2$ reduction of the Salam-Sezgin model, at least in the bosonic sector. It was observed that the bosonic Lagrangian of the Salam-Sezgin model can also be obtained from the noncritical string with the anomalous term in seven dimensions, which implies that the theory also contains a T-duality \cite{Ma:2023rdq,Ma:2023tcj}. The guaranteed consistent DeWitt $S^3$ reduction ensures the consistency of the $S^2$ Pauli reduction of the Salam-Sezgin model. Furthermore, this provides a simpler approach to construct the reduction ansatz than \cite{Gibbons:2003gp}. (See appendix \ref{app:3=2+1} for details.)

In this paper, we would like to generalize this relation between a supersymmetric theory and a non-supersymmetry one further. Utilising the warped reduction \cite{Kerimo:2004qx}, we find a curious duality in the bosonic sector of heterotic supergravity and noncritical bosonic string theory with conformal anomaly. The warped $\mathbb R$ reduction of heterotic supergravity with one vector multiplet yields the same nine-dimensional Lagrangian as the anomalous string on circle, but keeping the winding and Kaluza-Klein modes equal. This duality allows us to map the solutions of the bosonic string theory with $U(1)$ isometries to those of the heterotic theory. In particular, we are interested in the Mink$_4\times S^3\times S^3$ vacuum of the anomalous bosonic string. The internal compact $S^3\times S^3$ space can be viewed as an $U(1)$ bundle over $T^{1,1}$, where $T^{1,1}$ is a squashed non-Einstein metric on homogeneous coset of $[SU(2)\times SU(2)]/U(1)$. In other words, $S^3\times S^3 \sim S^1\ltimes T^{1,1}$. We can thus perform the duality on the $S^1$ and map the solution to be a warped geometry of $\mathbb R \times (T^{1,1}\times$Mink$_4)$ in the heterotic theory, supported by the heterotic NS-NS 3-form field strength, together with a Maxwell matter field strength. Note that the twisted fibre in $S^1\ltimes T^{1,1}$ is untwisted under the duality map and it becomes a warped product $\mathbb R\times T^{1,1}$. The solution can also be viewed as a vacuum in the heterotic theory since in its string frame, the geometry is homogeneous.

What is significant and highly non-obvious is that, although we obtain this unusual looking solution via a duality of low-energy effective theory between the supersymmetric heterotic theory and the non-supersymmetric bosonic string, we find that the warped $\mathbb R \times (T^{1,1}\times$Mink$_4)$ vacuum is supersymmetric, preserving $1/4$ of the heterotic supersymmetry. This suggests that we should have ${\cal N}=1$, $D=4$ supergravity arising from the Pauli-type of dimensional reduction of the heterotic theory on warped $\mathbb R\times T^{1,1}$ internal space. The reduction ansatz can be derived by starting with the simpler DeWitt type of $S^3\times S^3$ reduction of the anomalous bosonic string theory, with the left-invariant 1-forms kept intact.

The paper is organized as follows. In section 2, we derive the intriguing duality between the bosonic sector of heterotic supergravity and the anomalous bosonic string theory and establish the duality map between the two theories. We then derive the nontrivial warped $\mathbb R \times (T^{1,1}\times$Mink$_4)$ vacuum of the heterotic theory from the rather mundane Mink$_4\times S^3\times S^3$ vacuum of the bosonic string. We show that the new heterotic solution preserves $1/4$ of supersymmetry by constructing explicit Killing spinors. In section 3, we start from the simpler consistent $S^3\times S^3$ DeWitt reduction ansatz of the anomalous bosonic string theory and obtain the consistent warped $\mathbb R\times T^{1,1}$ reduction ansatz of the heterotic theory. We obtain the ${\cal N}=1$, $D=4$ supergravity with one complex scalar multiplet and $SU(2)\times SU(2)$ multiplets. In section 4, we present the reduction ansatz in a natural vielbein basis. We conclude the paper in section 5. In appendix \ref{app:formulae}, we collect many useful formulae associated with round $S^2$ and $S^3$. In appendix \ref{app:gend}, we extend the discussion in the main text to arbitrary dimensions. In appendix \ref{app:3=2+1}, we derive the $S^2$ reduction ansatz \cite{Gibbons:2003gp} of the Salam-Sezgin model from the relation between the bosonic sector of the model and the anomalous string.

\section{A duality between heterotic and bosonic strings}

In this section, we illustrate an intriguing duality between the bosonic sector of the ten-dimensional heterotic string theory and the anomalous bosonic string.
We observe that through a special warped $\mathbb R$ dimensional reduction \cite{Kerimo:2003am,Kerimo:2004qx}, the bosonic sector of heterotic string theory can be reduced to the nine-dimensional theory with positive exponential dilaton potential. Similarly, performing a standard $S^1$ Kaluza-Klein reduction on the noncritical bosonic string with conformal anomaly also leads to the same theory. This establishes a duality between the bosonic sectors of these two ten-dimensional theories. This duality allows us to map the solutions with $U(1)$ isometries of the anomalous bosonic string theory to the solutions in the heterotic theories. Intriguingly, some vacua of the bosonic string can yield nontrivial supersymmetric solutions in the heterotic theory.

\subsection{The bosonic duality}

We begin with the heterotic string theory, whose low-energy effective action is the ${\cal N}=1$, $D=10$ supergravity with either $E_8\times E_8$ or $SO(32)$ gauge fields. We shall only turn on the $U(1)$ gauge field. In other words, we consider ${\cal N}=1$, $D=10$ supergravity with one vector multiplet. The Lagrangian of the relevant bosonic sector is
\be
\hat{\mathcal{L}}_{10}= \hat{R}\, {\hat *\oneone}-\ft12 {\hat * d\hat \phi}\wedge d\hat \phi -\ft12 e^{-\hat\phi} {\hat *\hat H_\3} \wedge \hat H_\3 -\ft{1}{2 }e^{-\frac{1}{2}\hat{\phi}}{\hat *\hat F_\2}\wedge \hat F_\2\,,\label{heterotic}
\ee
where $\hat{F}_{\2}=d\hat{A}_{\1}$ and $\hat{H}_{\3}=d\hat{B}_{\2} +\frac{1}{2}\hat{F}_{\2}\wedge \hat{A}_{\1}$.
We consider the warped $\mathbb R$ reduction on the internal coordinate $z$
\bea
d\hat{s}_{10}^2&=&e^{2mz}\Big(
e^{-\frac{\bar{\phi}}{14a_9}}d\bar{s}_9^2+e^{\frac{\bar{\phi}}{2a_9}}dz^2
\Big),\qquad a_9=-\sqrt{\ft{8}{7}},\cr
\hat{\phi}&=&-\frac{1}{a_9}\bar{\phi}-4mz\,,\qquad \hat{B}_{\2}=\bar{B}_{\2}\,,\qquad \hat{A}_{\1}=\bar{A}_{\1}\,.\label{warped circle reduction}
\eea
Here, $m$ is an arbitrary constant. It can be verified that the reduction is consistent and we obtain the $D=9$ Lagrangian
\be
\bar{\mathcal{L}}_{9}=
\big(\bar{R} -64m^2e^{-\frac{1}{2}a_9\bar{\phi}}\big){\bar *\oneone}
-\ft12 {\bar *d\bar \phi}\wedge d\bar\phi -\ft12 e^{a_9\bar{\phi}}
{\bar * \bar H}_\3\wedge \bar H_\3
-\ft12 e^{\frac{1}{2}a_9\bar{\phi}} {\bar *\bar F}_\2\wedge \bar F_\2\,,\label{9D SS model}
\ee
where $\bar{H}_{\3}=d\bar{B}_{\2}+\frac{1}{2}\bar{F}_{\2}\wedge \bar{A}_{\1}$ and $\bar{F}_{\2}=d\bar{A}_{\1}$. This type of warped reduction \eqref{warped circle reduction} was first proposed in \cite{Kerimo:2003am}, and further developed in \cite{Kerimo:2004md,Kerimo:2004qx}. It is a semi-local gauging of the trombone symmetry \cite{Cremmer:1997xj} of supergravity and the constant shift symmetry of the dilaton scalar $\phi$, and it was sometimes named as the generalized Kaluza-Klein or Scherk-Schwarz reduction \cite{Lavrinenko:1997qa,
Bergshoeff:2002nv}. Here, ``semi-local gauging'' means that the constant transformation parameter of a global symmetry is taken to have a specific and appropriate dependence of the reduction coordinate. The resulting theory \eqref{9D SS model} resembles the low-energy effective action of the noncritical bosonic string theory, and it is indeed the case.

The Lagrangian of the noncritical bosonic string theory with conformal anomaly in ten dimensions takes the form
\bea
\check{\mathcal{L}}_{10}=
(\check{R}-8g^2 e^{\frac{1}{2}\check{\phi}})\,{\check * \oneone} -\ft12 {\check * d\check \phi}\wedge d\check\phi-\ft{1}{2}
e^{-\check{\phi}}{\check*\check{H}}_\3\wedge \check H_\3 \,,\qquad
\check{H}_{\3}=d\check{B}_{\2}\label{10D bosonic string}
\eea
We now perform the ordinary Kaluza-Klein $S^1$ reduction, but keeping the winding and Kaluza-Klein vectors equal. In other words, we consider the reduction ansatz
\bea
d\check{s}_{10}^2&=&e^{-\frac{\bar{\phi}}{14a_9}}d\bar{s}_9^2
+e^{\frac{\bar{\phi}}{2a_9}}\big(dz-\frac{1}{\sqrt{2}}\bar{A}_{\1}\big)^2,\cr
\check{\phi}&=&-\frac{1}{a_9}\bar{\phi}\,,\qquad \check{B}_{\2}=\bar{B}_{\2}+\frac{1}{\sqrt{2}}\bar{A}_{\1}\wedge dz\,.\label{normal circle reduction}
\eea
The reduction is again consistent and we recover the same $D=9$ Lagrangian \eqref{9D SS model}, provided that
\be
g^2=8m^2\,.\label{mgrelation}
\ee
Therefore, we have found a duality between the heterotic theory \eqref{heterotic} and the noncritical bosonic string with anomaly \eqref{10D bosonic string}. While the physical interpretation of this duality mapping can be questioned, it allows us to construct new solutions of heterotic theory \eqref{heterotic} of the warped type \eqref{warped circle reduction} from solutions in the anomalous string solution with an $U(1)$ isometry.

\subsection{The mapping of the vacua}

Having obtained the bosonic duality between the heterotic theory on \eqref{warped circle reduction} and the anomalous string on \eqref{normal circle reduction}, we can map the corresponding solutions that fit the reduction Ans\"atze from one theory to the other. It is clear that the warped solution of the type \eqref{warped circle reduction} in heterotic theory is uncommon, whilst solutions with an $U(1)$ isometry in the bosonic string theory is ubiquitous. One class of solutions that interest us is the Minkowski$\times$sphere vacua in the anomalous bosonic string theory. Owing the running exponential scalar potential associated with the conformal anomaly, the noncritical bosonic string theory does not have the Minkowski vacuum; instead, it admits Mink$_7\times S^3$ or Mink$_4\times S^3\times S^3$ vacua, supported by the magnetic 5-brane charges, together with the cosmological term.  We are interested in the latter case, since four-dimensional Minkowski spacetime emerges naturally:
\bea
d\check{s}_{10}^2&=&-dt^2 + dx^2 + dy^2 + dz^2+\frac{1}{g^2}\left(d\Omega_3^2+d\widetilde{\Omega}_3^2\right),\qquad \check{\phi}=0\,, \cr
\check{H}_{\3}&=&\frac{2}{g^2}\left(\omega_{\3}+\widetilde{\omega}_{\3}\right)
=\frac{1}{4g^2}\left(\sigma_1\wedge\sigma_2\wedge\sigma_3
+\tilde{\sigma}_{\tilde{1}}\wedge\tilde{\sigma}_{\tilde{2}}
\wedge\tilde{\sigma}_{\tilde{3}}\right).\label{D dim solution S3*S3}
\eea
To express the metrics of the two unit round $S^3$, we use two sets of $SU(2)$ left-invariant 1-forms $\sigma_i$, $\tilde{\sigma}_{\tilde{i}}$
\bea
&&\sigma_1=\cos\psi d\theta+\sin\psi\sin\theta d\varphi,\quad \sigma_2=-\sin\psi d\theta+\cos\psi\sin\theta d\varphi,\quad \sigma_3=d\psi+\cos\theta d\varphi,\cr
&&\tilde{\sigma}_{\tilde{1}}=\cos\tilde{\psi} d\tilde{\theta}+\sin\tilde{\psi}\sin\tilde{\theta} d\tilde{\varphi},
\quad \tilde{\sigma}_{\tilde{2}}=-\sin\tilde{\psi} d\tilde{\theta}+\cos\tilde{\psi}\sin\tilde{\theta} d\tilde{\varphi},\quad \tilde{\sigma}_{\tilde{3}}=d\tilde{\psi}+\cos\tilde{\theta} d\tilde{\varphi}\,.\label{sigma and the other sigma}
\eea
The metric and the volume form of $S^3$ are
\be
d\Omega_3^2=\ft{1}{4}\left(\sigma_1^2+\sigma_2^2+\sigma_3^2\right)\,,\qquad \omega_{\3}=\ft{1}{8}\sigma_1\wedge \sigma_2\wedge \sigma_3\,.
\ee
The same goes for the tilded version. We now rewrite the metric of $S^3\times S^3$ explicitly as
\bea
d\Omega_3^2+d\widetilde{\Omega}_3^2
&=&\ft{1}{8}\big(d\psi-d\tilde{\psi}+\cos\theta d\varphi-\cos\tilde{\theta} d\tilde{\varphi}\big)^2\cr
&&+\ft{1}{8}\big(d\psi+d\tilde{\psi}+\cos\theta d\varphi+\cos\tilde{\theta} d\tilde{\varphi}\big)^2
+\ft{1}{4}\big(d\Omega_2^2+d\widetilde{\Omega}_2^2\big).
\eea
This suggests to redefine the two $U(1)$ coordinates to be $\{\chi_1,\chi_2\}$, where
\bea
\psi+\tilde{\psi}=\chi_1\,,\qquad \psi-\tilde{\psi}=\chi_2\,,\qquad \Rightarrow\qquad \psi=\ft12(\chi_1+\chi_2)\,,\qquad \tilde{\psi}=\ft12(\chi_1-\chi_2)\,.\label{chi12}
\eea
Consequently, the $S^3\times S^3$ can be expressed as $S^1\ltimes T^{1,1}$, {\it i.e.} a $U(1)$ bundle over $T^{1,1}$:
\be
d\Omega_3^2+d\widetilde{\Omega}_3^2=\ft{1}{8}\big(d\chi_2+\cos\theta d\varphi-\cos\tilde{\theta} d\tilde{\varphi}\big)^2+\ft{1}{8}ds^2_{T^{1,1}}\,,
\ee
where
\be
ds^2_{T^{1,1}}=\big(d\chi_1+\cos\theta d\varphi+\cos\tilde{\theta} d\tilde{\varphi}\big)^2
+2\left(\sigma_1^2+\sigma_2^2+\tilde{\sigma}_{\tilde{1}}^2+
\tilde{\sigma}_{\tilde{2}}^2\right).\label{T11metric}
\ee
The $T^{1,1}$ is the coset space $[SU(2)\times SU(2)]/U(1)$ with a ``squashed'' homogeneous metric, where the term ``squashed'' signifies that the coefficient of the first term in \eqref{T11metric} has the nonstandard length of the $U(1)$ fibre associated with the Einstein metric of $T^{1,1}$.

In this set of new coordinates, the vacuum solution \eqref{D dim solution S3*S3} can now be expressed as
\bea
d\check{s}_{10}^2&=&d\bar{s}_{9}^2+\frac{1}{8g^2}\left(d\chi_2+\cos\theta d\varphi-\cos\tilde{\theta} d\tilde{\varphi}\right)^2\,,\qquad
d\bar{s}_{9}^2=\eta_{\mu\nu}dx^\mu dx^\nu+\frac{1}{8g^2}ds^2_{T^{1,1}}\,,\cr
\check{H}_{\3}
&=&\frac{1}{8g^2}\left(\sigma_1\wedge\sigma_2-\tilde{\sigma}_{\tilde{1}}
\wedge\tilde{\sigma}_{\tilde{2}}\right)\wedge d\chi_2+\frac{1}{8g^2}\left(\sigma_1\wedge\sigma_2+
\tilde{\sigma}_{\tilde{1}}\wedge\tilde{\sigma}_{\tilde{2}}
\right)\wedge d\chi_1\,.
\eea
Here $\eta_{\mu\nu}dx^\mu dx^\nu$ denotes the four-dimensional Minkowski spacetime.

By defining the coordinate $\chi_2=2\sqrt2 g\, z$, we find that the solution fits the reduction ansatz \eqref{normal circle reduction}, which leads to a nine-dimensional solution of the theory \eqref{9D SS model}. Specifically, we have
\bea
d\bar{s}_9^2&=&\eta_{\mu\nu}dx^\mu dx^\nu+\frac{1}{64m^2}ds^2_{T^{1,1}},\quad \bar{\phi}=0\cr
\bar{B}_{\2}&=&-\frac{1}{64m^2}\left(\cos\theta d\varphi+\cos\tilde{\theta} d\tilde{\varphi}
\right)\wedge d\chi_1,\quad \bar{A}_{\1}=-\frac{\sqrt{2}}{8m}\left(\cos\theta d\varphi-\cos\tilde{\theta} d\tilde{\varphi}
\right),
\eea
where we have converted to the parameter $g$ to $m$ by \eqref{mgrelation}. We can now lift the solution back to the ten dimensions via the reduction ansatz \eqref{warped circle reduction}. We thus obtain the vacuum solution of the $D=10$ heterotic theory
\bea
d\hat{s}_{10}^2&=&e^{2mz}\Big(\eta_{\mu\nu}dx^\mu dx^\nu+\frac{1}{64m^2}ds^2_{T^{1,1}}+dz^2
\Big),\qquad \hat{\phi}=-4mz, \cr
\hat{B}_{\2}&=&-\frac{1}{64m^2}\big(\cos\theta d\varphi+\cos\tilde{\theta} d\tilde{\varphi} \big)\wedge d\chi_1\,,\qquad \hat{A}_{\1}=-\frac{\sqrt{2}}{8m}\big(\cos\theta d\varphi-\cos\tilde{\theta} d\tilde{\varphi}\big).\label{hetsol}
\eea
It can be independently verified that the above solution indeed satisfies all the equations of motion of \eqref{heterotic}. Note that we call the solution a string vacuum in that in the string frame, namely
\be
d\hat s_{\rm str}^2 = e^{\fft12\hat\phi} d\hat s_{\rm Ein}^2=\eta_{\mu\nu}dx^\mu dx^\nu+\frac{1}{64m^2}ds^2_{T^{1,1}}+dz^2\,,
\ee
the metric is homogeneous. We can also use the 1-forms \eqref{sigma and the other sigma} to express the form field strengths in the heterotic theory
\bea
\hat{F}_{\2}&=&\frac{\sqrt{2}}{8m}\left(\sigma_1\wedge\sigma_2-
\tilde{\sigma}_{\tilde{1}}\wedge\tilde{\sigma}_{\tilde{2}}
\right),\cr
\hat{H}_{\3}&=&\frac{1}{64m^2}\left(\sigma_1\wedge\sigma_2+
\tilde{\sigma}_{\tilde{1}}\wedge\tilde{\sigma}_{\tilde{2}}
\right)\wedge \big(d\chi_1+\cos\theta d\varphi+\cos\tilde{\theta} d\tilde{\varphi}\big).\label{hetfs}
\eea
It can be checked that the 3-form satisfies the Bianchi identity
\bea
d\hat{H}_{\3}=\ft{1}{2}\hat{F}_{\2}\wedge\hat{F}_{\2}.
\eea

\subsection{Supersymmetry}

We obtained a string vacuum in the heterotic theory by performing the bosonic duality between the heterotic theory and the anomalous bosonic string. This is in itself significant since \eqref{hetsol} is not the most immediate Ans\"atze, while Mink$_4\times S^3\times S^3$ is rather natural in the anomalous bosonic string. However, it will be more significant if the solution turns out to be supersymmetric.

At first sight, the idea of supersymmetry appears to be farfetched, since the anomalous bosonic string is not supersymetric. However, although the bosonic string is not supersymmetric, it is pseudo-supersymmetric \cite{Lu:2011nga,Lu:2011zx}, even when the conformal anomaly term \cite{Lu:2011ku} or the $\alpha'$ corrections \cite{Lu:2011vk}
are included. Pseudo-supersymmetry means that up to and including the quadratic orders of the fermion fields, the Lagrangian is invariant by the pseudo-supersymmetry transformation rule \cite{Lu:2011nga,Lu:2011zx, Hassler:2020rlq}. This is precisely the order where Killing spinors are defined, and furthermore, the integrability condition of the Killing spinor equations gives rise precisely to the bosonic equations of motion. It is easy to establish that Mink$_4\times S^3\times S^3$ has Killing spinors under pseudo-supersymmetric transformation rule.

However, as we have mentioned that, Mink$_7\times S^3$ is also a solution of the anomalous bosonic string and it also admits Killing spinors. We can map this solution by the duality transformation to become a warped metric of $\mathbb R \times (S^2 \times$Mink$_7)$ in the heterotic theory, with the matter $U(1)$ field strength $F_\2$ carrying magnetic $S^2$ flux. We can immediately deduce the warped metric of $\mathbb R \times (S^2 \times$Mink$_7)$ in the heterotic theory is not supersymmetric, since the Killing spinor equation of the heterotic string requires
\be
F_{MN} \Gamma^{MN} \epsilon=0\,.\label{kseq0}
\ee
This is not possible when the matter $F_\2\propto \sigma_1\wedge\sigma_2$. This however does not necessarily imply that the pseudo-supersymmetry has no relevance in supersymmetry, since it is not inconsistent with the fact that Mink$_7\times S^3$ is pseudo-supersymmetric in the anomalous string side. The Killing spinor in Mink$_7\times S^3$ can be decomposed into the $S^3$ direction and it is given by
\be
\epsilon \sim e^{\pm {\rm i} \psi}\epsilon_0\,.
\ee
(See Killing spinor constructions in $S^n$ \cite{Lu:1998nu}.) Thus it is the reduction on the $U(1)$ fibre coordinate $\psi$ that breaks the pseudo-supersymmetry, and hence the supersymmetry.

The situation for Mink$_4\times S^3\times S^3$ changes, and the Killing spinors take the form
\be
\epsilon \sim e^{\pm {\rm i} \psi \pm {\rm i} \tilde\psi}\epsilon_0\,.
\ee
In the Kaluza-Klein $S^1$ reduction on $\chi_2$, defined by \eqref{chi12}, it is clear that half of the Kiling spinors will be retained. Intriguingly, The solution of $\hat F_\2$ given in \eqref{hetfs} implies that the Killing spinor equation \eqref{kseq0} will also project out half of the spinors. These arguments suggest that the heterotic solution may be supersymmetric, but the ultimate test is to construct the Killing spinors explicitly.

   The fermionic sector for the heterotic supergravity consists of the
gravitino, dilatino and gaugino, and the corresponding Killing spinor equations are (e.g.~\cite{Lu:2006ah})
\bea
\delta \hat \psi_M &=& \hat D_M \hat \epsilon - \fft1{96} e^{-\fft12\hat \phi}\big(
\Gamma_M \Gamma^{PQR} - 12 \delta_M^P \Gamma^{QR}\big) \hat H_{PQR}\, \hat \epsilon=0\,,\nn\\
\delta \hat \chi &=& -\fft12 \Gamma^M \partial_M\phi\,\hat \epsilon -
\fft12 e^{-\fft12 \hat \phi} \Gamma^{MNP} \hat H_{MNP}\,\hat \epsilon=0\,,\nn\\
\delta \hat \lambda &=& -\fft12 e^{-\fft14 \hat\phi} \Gamma^{MN} \hat F_{MN}\, \hat \epsilon =0\,.\label{susy}
\eea
The Killing spinors of the heterotic supergravity are chiral and satisfy the projection
\be
\Gamma^{11} \hat \epsilon=\hat \epsilon\,.\label{chiral}
\ee
To calculate the Killing spinors, we adopt the following vielbein choice:
\bea
&&e^{\underline\mu}=e^{m z} dx^\mu\,,\qquad e^4= \fft{\sqrt2 e^{mz}}{8m}\,d\theta_1\,,\qquad
e^5=\fft{\sqrt{2} e^{mz}}{8m}\sin\theta_1\,d\phi_1\,,\nn\\
&&e^6= \fft{\sqrt2 e^{mz}}{8m}\,d\theta_2\,,\qquad
e^7=\fft{\sqrt{2} e^{mz}}{8m} \sin\theta_2\,d\phi_2\,,\nn\\
&&e^8= \fft{e^{mz}}{8m} (d\chi_1 + \cos\theta_1\, d\phi_1 + \cos\theta_2\, d\phi_2)\,,\qquad
e^9 = e^{m z} dz\,.
\eea
After a careful calculation, we find that the Killing spinors take the form
\be
\epsilon = e^{\fft12 m z}\,\Big(e^{\fft{\rm i}{2} \chi_1}\epsilon_+ + e^{-\fft{\rm i}{2} \chi_1}\epsilon_-\Big)\,,
\ee
where $\epsilon_\pm$ are constant spinors satisfying the projection
\be
\Gamma^{45}\epsilon_\pm = \Gamma^{67}\epsilon_\pm=\Gamma^{89} \epsilon_\pm
=\mp {\rm i}\, \epsilon_\pm\,.
\ee
Note that these projection condition is consistent with the chiral projection \eqref{chiral}. Therefore, the heterotic solution is indeed supersymmetric and it preserves $1/4$ of the heterotic supersymmetry. We thus have constructed a concrete example of supersymmetric vacuum of heterotic supergravity, {\it via} a pseudo-supersymmetric vacuum of the bosonic string with conformal anomaly.

\section{Warped $\mathbb R \times T^{1,1}$ reduction of heterotic theory}

In the previous section, we obtained a supersymmetric heterotic string vacua that preserves one quarter of the supersymmetry, {\it via} the rather unusual bosonic duality between heterotic supergravity and the anomalous bosonic string. This suggests that we can obtain an ${\cal N}=1$ $D=4$ supergravity with Minkowski vacuum if we reduce heterotic supergravity on this background. We now carry out the reduction in this section and obtain four-dimensional supergravity with $SU(2)\times SU(2)$ Yang-Mills, where the origin of the gauge symmetry is the isometry of the $T^{1,1}$ space.

Consistent Kaluza-Klein reduction on non-maximally-symmetric space such as $T^{1,1}$ is subtle. For example, it was shown to be inconsistent to perform Pauli-type reduction of type IIB supergravity on the Einstein metric of $T^{1,1}$ \cite{Hoxha:2000jf}, even though the AdS$_5\times T^{1,1}$ is its vacuum. (However, there does exist a supersymmetric consistent truncation where only the $U(1)$ gauge field is retained \cite{Cassani:2010na,Bena:2010pr}.) The situation here is different, and we again take the advantage of the bosonic duality between the heterotic supergravity and the anomalous string. We first perform the $S^3\times S^3$ reduction of the anomalous string and then map the reduction ansatz to that of the heterotic theory on the warped $\mathbb R\ltimes T^{1,1}$. It should be emphasized that without this unusual duality between the two theories, we would not have dreamt to consider the possibility of the $T^{1,1}$ Pauli reduction, based on the work of \cite{Hoxha:2000jf}.

\subsection{Consistent $S^3\times S^3$ reduction of $D=10$ anomalous string}

The bosonic string theory reduced on $S^3$ \cite{Cvetic:2000dm} or more generally group manifolds \cite{Lu:2006ah} has been previously discussed. Together with the Mink$_4\times S^3\times S^3$ vacuum obtained in the previous section, we find that the DeWitt type of reduction ansatz for $S^3 \times S^3$ is
\bea
d\check{s}_{10}^2&=&e^{\frac{3}{4}\phi}ds_{4}^2
+\frac{1}{4g^2}e^{-\fft14\phi}\left(h^\alpha h_\alpha+\tilde{h}^{\tilde{\alpha}} \tilde{h}_{\tilde{\alpha}}\right),\label{s3s3red}
\eea
where
\be
h^\alpha=\sigma^\alpha - g A_{\1}^\alpha\,,\qquad \tilde h^{\tilde\alpha} =\sigma^{\tilde\alpha} - g \widetilde A_{\1}^{\tilde \alpha}\,.\label{hdef}
\ee
The reduction of the dilaton is simply $\check{\phi}=\fft12\phi$. The ansatz for $\check{B}_{\2}$ is more involved, given by
\bea
\check{B}_{\2}&=&B_{\2}+\frac{1}{4g}(\cos\psi A_{\1}^1-\sin\psi A_{\1}^2)\wedge d\theta+\frac{1}{4g}A_{\1}^3\wedge d\psi\cr
&&+\frac{1}{4g}(\cos\tilde{\psi} \widetilde{A}_{\1}^{\tilde{1}}-\sin\tilde{\psi} \widetilde{A}_{\1}^{\tilde{2}})\wedge d\tilde{\theta}+\frac{1}{4g}\widetilde{A}_{\1}^{\tilde{3}}\wedge d\tilde{\psi}
\cr
&&-\frac{1}{4g^2}\left(\cos\theta d\psi-g\mu_\alpha A_{\1}^\alpha\right)\wedge d\varphi -\frac{1}{4g^2}\big(\cos\tilde{\theta} d\tilde{\psi}-g\tilde{\mu}_{\tilde{\alpha}} \widetilde{A}_{\1}^{\tilde{\alpha}}\big)\wedge d\tilde{\varphi}\,.
\label{s3s3-2forms}
\eea
The coordinates $\mu^\alpha$ and $\tilde \mu^{\tilde \alpha}$ will be explained later and also in appendix \ref{app:formulae}. The $SU(2)$ group indices $\alpha$ and $\tilde\alpha$ are Euclidean and we are not strict with their superscript or subscript positions. Under these reduction Ans\"atze, we find that the $D=10$ anomalous string theory \eqref{10D bosonic string} can be consistently reduced to the four-dimensional theory with the Lagrangian
\be
\mathcal{L}_{4}=R\, {*\oneone}-\ft{1}{2} {* d\phi}\wedge d\phi-\ft{1}{2}e^{-2\phi}{*H_{\3}}\wedge H_{\3}
-\ft{1}{4}e^{-\phi}\big({*F_{\2}^\alpha}\wedge F_{\2\alpha}
+{*\widetilde{F}_{\2}^{\tilde{\alpha}}}\wedge \widetilde{F}_{\2\tilde{\alpha}}\big),\label{d4theory}
\ee
where the 3-form and 2-form field strengths are
\bea
H_{\3}&=&dB_{\2}+\ft{1}{4}\big(F_{\2}^\alpha\wedge A_{\1\alpha}+\widetilde{F}_{\2}^{\tilde{\alpha}}\wedge \widetilde{A}_{\1{\tilde{\alpha}}}\big)\cr
&&-\ft{1}{24}g\big(\epsilon_{\alpha\beta\gamma}A_{\1}^\alpha\wedge A_{\1}^\beta\wedge A_{\1}^\gamma+\epsilon_{\tilde{\alpha}\tilde{\beta}\tilde{\gamma}}
\widetilde{A}_{\1}^{\tilde{\alpha}}\wedge \widetilde{A}_{\1}^{\tilde{\beta}}\wedge \widetilde{A}_{\1}^{\tilde{\gamma}}\big),\cr
F_{\2}^\alpha&=&dA_{\1}^\alpha+\ft{1}{2}g\epsilon^{\alpha\beta\gamma}
A_{\1\beta}\wedge A_{\1\gamma}\,,\qquad
\widetilde{F}_{\2}^{\tilde{\alpha}}=d\widetilde{A}_{\1}^{\tilde{\alpha}}
+\ft{1}{2}g\epsilon^{\tilde{\alpha}\tilde{\beta}\tilde{\gamma}}
\widetilde{A}_{\1\tilde{\beta}}\wedge \widetilde{A}_{\1\tilde{\gamma}}\,.\label{h3f2}
\eea
The 3-form field strength $H_\3$ satisfies the Bianchi identity
\be
dH_{\3}=\ft{1}{4}\big(F_{\2}^\alpha\wedge F_{\2\alpha}+\widetilde{F}_{\2}^{\tilde{\alpha}}
\wedge \widetilde{F}_{\2\tilde{\alpha}}\big)\,.\label{h3bianchi}
\ee
(Note that there is an extra $1/2$ factor in both the Bianchi identity and the kinetic terms of the Yang-Mills fields in our convention.) This 3-form field strength in four dimensions is Hodge dual of a 1-form, associated with an axion field $\chi$:
\be
e^{-2\phi}{*H_{\3}}=d\chi\,.
\ee
The four-dimensional Lagrangian then becomes
\bea
\mathcal{L}_{4}&=&R\, {*\oneone}-\ft12 {* d\phi}\wedge d\phi-\ft12e^{2\phi}{* d\chi}\wedge d\chi
-\ft14e^{-\phi}\big({*F_{\2}^\alpha}\wedge F_{\2\alpha}
+{* \widetilde{F}_{\2}^{\tilde{\alpha}}}\wedge \widetilde{F}_{\2\tilde{\alpha}}\big)\nn\\
&&+ \ft14\chi \big(F_{\2}^\alpha\wedge F_{\2\alpha}
+\widetilde{F}_{\2}^{\tilde{\alpha}}\wedge \widetilde{F}_{\2\tilde{\alpha}}\big).
\eea
Thus we arrive at four-dimensional Einstein gravity coupled to $SU(2)\times SU(2)$ Yang-Mills fields, together with a complex scalar $\chi + {\rm i} e^{\phi}$ that forms a scalar coset of $SL(2,\mathbb R)/SO(2)$. Next, we show that this same theory can also be obtained from warped $\mathbb R\times T^{1,1}$ reduction of the heterotic theory.

\subsection{Reduction ansatz on $T^{1,1}$}

As we have discussed, the internal $S^3\times S^3$ space in the previous subsection can be viewed also as $S^1\ltimes T^{1,1}$. The reduction on the $S^1$ {\it via} \eqref{normal circle reduction} yields a nine-dimensional theory \eqref{9D SS model}. We can thus derive the $T^{1,1}$ reduction ansatz of \eqref{9D SS model} to obtain the four-dimensional theory. Specifically, we note that in the round $S^3$ metric $\fft14\sigma^\alpha\sigma^\alpha$, there are two commuting $U(1)$ directions associated with coordinates $(\psi, \varphi)$. However, when the Yang-Mills fields are turned on, the metric $\fft14h^\alpha h^\alpha$ has only one $U(1)$ isometry, associated with coordinates $\varphi$. The same goes with the tilded $S^3$.

We would like to reduce the ten-dimensional $S^3\times S^3$ on the diagonal $S^1$ first. We thus define the orthogonal combination $\{z,\tau\}$ of the two $U(1)$ coordinates $\{\varphi,\tilde{\varphi}\}$ in $S^3\times S^3$ by
\bea
\varphi-\tilde{\varphi}=2\sqrt{2}gz\,,\qquad \varphi+\tilde{\varphi}=\tau\,.\label{ztau}
\eea
Comparing to the circle reduction ansatz \eqref{normal circle reduction} with $S^3\times S^3$ reduction ansatz given in the previous subsection, we obtain the $T^{1,1}$ reduction ansatz of the $D=9$ Lagrangian \eqref{9D SS model}. Specifically, we have
\bea
d\bar{s}_9^2&=&e^{\frac{5}{7}\phi}ds_{4}^2 +\frac{1}{g^2}e^{-\frac{2}{7}\phi}\left[\ft{1}{8}\big(\Sigma-g\mu_\alpha A_{\1}^\alpha -g\tilde{\mu}_{\tilde{\alpha}} \widetilde{A}_{\1}^{\tilde{\alpha}}
\big)^2+\ft{1}{4}\left(D\mu^\alpha D\mu_\alpha+D\tilde{\mu}^{\tilde{\alpha}} D\tilde{\mu}_{\tilde{\alpha}}\right)
\right],\cr
\bar{B}_{\2}&=&B_{\2}+\frac{1}{4g}(\cos\psi A_{\1}^1-\sin\psi A_{\1}^2)\wedge d\theta+\frac{1}{4g}A_{\1}^3\wedge d\psi\cr
&&+\frac{1}{4g}(\cos\tilde{\psi} \widetilde{A}_{\1}^{\tilde{1}}-\sin\tilde{\psi} \widetilde{A}_{\1}^{\tilde{2}})\wedge d\tilde{\theta}+\frac{1}{4g}\widetilde{A}_{\1}^{\tilde{3}}\wedge d\tilde{\psi}\cr
&&-\frac{1}{8g^2}\left(\Sigma-g\mu_\alpha A_{\1}^\alpha
-g\tilde{\mu}_{\tilde{\alpha}} \widetilde{A}_{\1}^{\tilde{\alpha}}
\right)\wedge d\tau,\qquad \Sigma=d\tau+\cos\theta d\psi+\cos\tilde{\theta} d\tilde{\psi}\,,\cr
\bar{A}_{\1}&=&\frac{1}{2g}\left(\cos\tilde{\theta} d\tilde{\psi}-\cos\theta d\psi
-g\tilde{\mu}_{\tilde{\alpha}} \widetilde{A}_{\1}^{\tilde{\alpha}}+g\mu_\alpha A_{\1}^\alpha
\right),\qquad
\bar{\phi}=\sqrt{\fft27}\phi\,.\label{T11 ansatz}
\eea
Although it is a nontrivial exercise, it can be verified that the above ansatz can be used to consistently reduce the theory \eqref{9D SS model} to the four-dimensional one given by \eqref{d4theory}.

\subsection{Warped $\mathbb R\times T^{1,1}$ reduction of heterotic theory}

Once we have the consistent $T^{1,1}$ reduction ansatz of the nine-dimensional theory \eqref{9D SS model}, the reduction ansatz of the heterotic theory \eqref{heterotic} follows straightforwardly from \eqref{warped circle reduction}, namely
\bea
d\hat s_{10}^2 &=& e^{2mz}\Bigg(e^{\fft34 \phi} ds_4^2
+e^{-\frac{1}{4}\phi}\Big[ dz^2\nn\\
&&+\frac{1}{8g^2}\left(\Sigma-g\mu_\alpha A_{\1}^\alpha -g\tilde{\mu}_{\tilde{\alpha}} \widetilde{A}_{\1}^{\tilde{\alpha}}\right)^2+\frac{1}{4g^2}\left(D\mu^\alpha D\mu_\alpha+D\tilde{\mu}^{\tilde{\alpha}} D\tilde{\mu}_{\tilde{\alpha}}\right)\Big]\Bigg)\,,\nn\\
\hat B_{\2} &=& \bar B_{\2}\,,\qquad \hat A_{\1} = \bar A_{\1}\,,\qquad
\hat \phi = \ft12 \phi - 4mz\,,
\eea
together with the $m$ and $g$ relation \eqref{mgrelation}.

We thus obtain supergravity with $SU(2)\times SU(2)$ Yang-Mills fields in
four dimensions. The supersymmetry can be established to be ${\cal N}=1$. The vacuum of the theory is Minkowski. One salient point is that the Yang-Mills fields all have geometric origin associated with the isometry group of the internal $T^{1,1}$ space. An analogous theory in literature is the ${\cal N}=1$, $D=4$ supergravity that arises from the $S^2$ reduction \cite{Gibbons:2003gp} of the Salam-Sezgin model \cite{Salam:1984cj}, which we shall discuss further in appendix \ref{app:3=2+1}. The whole discussion in this section from ten to four dimensions can be generalized to arbitrary $D$ to $D-6$ dimensions, given in appendix \ref{app:gend}.

\section{Reduction in a vielbein basis}

In the previous section, we obtained the nontrivial and consistent Pauli-type reduction ansatz of the heterotic theory on warped $\mathbb R\times T^{1,1}$ space to obtain the four-dimensional supergravity \eqref{d4theory}. The reduction was performed in the coordinate base and it was on the gauge field potentials rather than on their field strengths. For the purpose of future analysing supersymmetry, it is advantageous to obtain the reduction ansatz for field strengths in the vielbein basis.

The metric ansatz \eqref{s3s3red} is already in the natural diagonal base. The corresponding natural choice vielbein is obvious and we shall not present in detail. The reduction ansatz for the 3-form field strength can be obtained from its 2-form gauge potential \eqref{s3s3-2forms}. We find
\be
\check{H}_{\3}=H_{\3}-\frac{1}{4g^2}\big(\Omega_{\3}+\widetilde{\Omega}_{\3}
\big)+\frac{1}{4g}F_{\2}^\alpha\wedge h_\alpha+\frac{1}{4g}\widetilde{F}_{\2}^{\tilde{\alpha}}\wedge \tilde{h}_{\tilde{\alpha}}\,,\label{checkH3}
\ee
where
\be
\Omega_{\3}=\ft{1}{6}\epsilon_{\alpha\beta\gamma}h^\alpha\wedge h^\beta\wedge h^\gamma\,,\qquad
\widetilde{\Omega}_{\3}=\ft{1}{6}\epsilon_{\tilde{\alpha}\tilde{\beta}
\tilde{\gamma}}h^{\tilde{\alpha}}\wedge h^{\tilde{\beta}}\wedge h^{\tilde{\gamma}}.
\ee
The four-dimensional field strengths $H_\3$,  $F_\2^\alpha$ and
$\widetilde F_\2^{\tilde\alpha}$ were given in \eqref{h3f2}. The 10-dimensional Bianchi identity $d\hat H_{\3}=0$ yields the 4-dimensional Bianchi identity \eqref{h3bianchi}.

In order to extract the diagonal $U(1)$ direction of $S^3\times S^3$ in the vielbein basis, we present many useful formulae associated with the properties of $S^2$ and $S^3$ in appendix \ref{app:formulae}. As explained in the appendix, we have
\bea
h^\alpha h_\alpha&=&D\mu^\alpha D\mu_\alpha+\sigma^2\,,\qquad \sigma=d\varphi+\mathcal{A}_{\1}\,,\qquad \mathcal{A}_{\1} =\cos\theta d\psi-g\mu_\alpha A_{\1}^\alpha\,,\cr
\tilde{h}^{\tilde{\alpha}} \tilde{h}_{\tilde{\alpha}}&=&D\tilde{\mu}^{\tilde{\alpha}} D\tilde{\mu}_{\tilde{\alpha}}+\tilde{\sigma}^2\,,\qquad \tilde{\sigma}=d\tilde{\varphi}+\widetilde{\mathcal{A}}_{\1}\,,\qquad
\widetilde{\mathcal{A}}_{\1}=\cos\tilde\theta d\tilde\psi-g\tilde\mu_{\tilde\alpha} \widetilde A_{\1}^{\tilde\alpha}\,.
\eea
We can thus again define the coordinates $\{z, \tau\}$ as in \eqref{ztau}.
Therefore, the $S^3 \times S^3 $ in the metric ansatz \eqref{s3s3red} can be written as $S^1 \ltimes T^{1,1}$, even with the $SU(2)\times SU(2)$ Yang-Mills turned on:
\be
\ft{1}{4}\big(h^\alpha h_\alpha+\tilde{h}^{\tilde{\alpha}} \tilde{h}_{\tilde{\alpha}}\big)
=g^2\, \Sigma_2^2 +\ft{1}{8}\,\Sigma_1^2 +\ft{1}{4} \left(
D\mu^\alpha D\mu_\alpha+D\tilde{\mu}^{\tilde{\alpha}} D\tilde{\mu}_{\tilde{\alpha}}
\right),
\ee
where
\be
\Sigma_1 =d\tau+\mathcal{A}_{\1}+\widetilde{\mathcal{A}}_{\1}\,,\qquad
\Sigma_2 = dz+\frac{1}{2\sqrt{2}g}(\mathcal{A}_{\1}-\widetilde{\mathcal{A}}_{\1})\,.
\ee
Performing the $S^1$ reduction on $z$ coordinate, (or on the vielbein $\Sigma_2$,) we obtain the $D=9$ metric
\bea
d\bar{s}_{9}^2&=&e^{\frac{5}{7}\phi}ds_4^2 +\frac{1}{g^2}e^{-\frac{2}{7}\phi}\Big(\ft{1}{8}\Sigma_1^2
+\ft{1}{4}(\varepsilon^a\varepsilon^a+\tilde{\varepsilon}^{\tilde{a}}
\tilde{\varepsilon}^{\tilde{a}})
\Big).
\eea
Note that it follows from the useful formulae in appendix \ref{app:formulae}, we have
\be
D\mu^\alpha D\mu_\alpha=\varepsilon^a\varepsilon^a\,,\qquad
D\tilde{\mu}^{\tilde{\alpha}} D\tilde{\mu}_{\tilde{\alpha}}=\tilde{\varepsilon}^{\tilde{a}}
\tilde{\varepsilon}^{\tilde{a}}\,.
\ee
(See the definition of $\varepsilon$ in \eqref{varep}.) To derive the ansatz for the field strengths in the vielbein basis, we write $\Omega_{\3}$ and $h^\alpha$ as
\bea
\Omega_{\3}&=&\frac{1}{2}\epsilon_{\alpha\beta\gamma}\mu^\alpha D\mu^\beta\wedge D\mu^\gamma\wedge\sigma=-\frac12 \epsilon_{ab}\, \varepsilon^a\wedge\varepsilon^b\wedge \sigma\,,\cr
h^\alpha &=& -\epsilon^{\alpha\beta\gamma}\mu_\beta D\mu_\gamma+\mu^\alpha\sigma = K_\alpha^a \varepsilon^a + \mu^\alpha \sigma\,.
\eea
Analogous expressions hold for the tilded variables. The 3-form field strength \eqref{checkH3} becomes
\bea
\check{H}_{\3}&=&H_{\3}+\frac{1}{4g}K_\alpha^a F_{\2}^\alpha\wedge \varepsilon^a
+\frac{1}{4g}\widetilde{K}_{\tilde{\alpha}}^{\tilde{a}}
\widetilde F_{\2}^{\tilde{\alpha}}\wedge\tilde{\varepsilon}^{\tilde{a}}\cr
&&+\fft{1}{8g^2}\big(2g\mu_\alpha F_{\2}^\alpha+\epsilon_{ab}\,\varepsilon^a
\wedge\varepsilon^b\big)\wedge \sigma +\fft{1}{8g^2}\big(2g\tilde{\mu}_{\tilde{\alpha}} \widetilde F_{\2}^{\tilde{\alpha}}
+\epsilon_{\tilde{a}\tilde{b}}\,
\tilde{\varepsilon}^{\tilde{a}}\wedge\tilde{\varepsilon}^{\tilde{b}}\big)
\wedge\tilde{\sigma}\,.
\eea
We can thus read off the reduction ansatz of $\bar{H}_{\3}$ in nine dimensions:
\bea
\bar{H}_{\3}&=&H_{\3}+\frac{1}{4g}K_\alpha^a F_{\2}^\alpha\wedge \varepsilon^a
+\frac{1}{4g}\widetilde{K}_{\tilde{\alpha}}^{\tilde{a}}
F_{\2}^{\tilde{\alpha}}\wedge \tilde{\varepsilon}^{\tilde{a}}\cr
&&+\frac{1}{16g^2}\Big(2g(\mu_\alpha F_{\2}^\alpha + \tilde{\mu}_{\tilde{\alpha}} \widetilde F_{\2}^{\tilde{\alpha}})+\epsilon_{ab}\, \varepsilon^a\wedge \varepsilon^b + \epsilon_{\tilde{a}\tilde{b}}\, \tilde\varepsilon^{\tilde a}\wedge \tilde \varepsilon^{\tilde b}\Big) \wedge\Sigma_1\,.
\eea
Similarly, we can also express \(\bar{F}_{\2}\) in the vielbein basis:
\bea
\bar{F}_{\2}=\ft{1}{2}(\mu_\alpha F_{\2}^\alpha-\tilde{\mu}_{\tilde{\alpha}} \widetilde F_{\2}^{\tilde{\alpha}})
+\fft{1}{4g}\Big( \epsilon_{ab}\, \varepsilon^a\wedge \varepsilon^b
-\epsilon_{\tilde{a}\tilde{b}}\, \tilde \varepsilon^{\tilde a}\wedge \tilde \varepsilon^{\tilde b}\Big).
\eea

With these preliminaries, we are ready to give the warped $\mathbb R\times T^{1,1}$ reduction in the vielbein basis
\bea
d\hat{s}_{10}^2&=&e^{2mz}\left(
e^{\fft34\phi}ds_4^2+ e^{-\fft14\phi} \Big(\frac{1}{8g^2}\Sigma_1^2
+\frac{1}{4g^2}(\varepsilon^a\varepsilon^a+\tilde{\varepsilon}^{\tilde{a}}
\tilde{\varepsilon}^{\tilde{a}})+ dz^2\Big)\right),\cr
\hat{H}_{\3}&=& \bar H_\3\,,\qquad \hat{F}_{\2}=\bar F_\2\,,\qquad
\hat{\phi}=\ft12 \phi-4mz,\label{warped circle T11 reduction}
\eea
This reduction ansatz allows one to work out the reduction ansatz on the fermions and obtain the four-dimensional supersymmetric transformation rules from the ten-dimensional ones given in \eqref{susy}. In a forthcoming paper, C.N.~Pope is constructing the fermionic reduction in this same model.

\section{Conclusions}

In this paper, we observed a curious duality between the bosonic sector of the heterotic supergravity and the noncritical bosonic string with the conformal anomaly term. The explicit duality map allows us to obtain a new vacuum of warped $\mathbb R\times (T^{1,1}\times$Mink$_4)$ geometry in heterotic theory, supported by both the magnetic fluxes of the 3-form and the matter Maxwell 2-form field strengths. Remarkably, although our starting point of the bosonic string is non-supersymmetric, the derived heterotic vacuum is supersymmetric, preserving $1/4$ of the supersymmetry. We verified this by constructing the explicit Killing spinors.

Our way of constructing new supergravity vacua from a non-supersymmetric starting point may not be a coincidence, since the bosonic string theory has pseudo-supersymmetry. By pseudo-supersymmetry, it means that theory admits Killing spinor equations, whose integrability condition gives precisely the bosonic equations of motion \cite{Lu:2011nga,Lu:2011zx}. The simplest such example is Einstein gravity in general dimensions. A consequence of the pseudo-supersymmetry is that we can construct a Lagrangian of both the bosonic and fermionic sectors, and it is invariant under the pseudo-supersymmetric transformation, up to and including the quadratic orders of fermions. It turns out that the bosonic string theory, including the anomaly term \cite{Lu:2011ku} and $\alpha'$ corrections \cite{Lu:2011zx}, is also pseudo-supersymmetric. Thus, it is the pseudo-supersymmetric Mink$_4\times S^3\times S^3$ vacuum of the bosonic string that is mapped to the supersymmetric warped $\mathbb R\times (T^{1,1}\times$Mink$_4)$ in heterotic theory. Although pseudo-supersymmetry has been largely ignored in literature, our work gave a nontrivial example of its role in constructing supersymmetric solutions in supergravities.

We obtained the bosonic sector of ${\cal N}=1$, $D=4$ supergravity with one complex scalar and $SU(2)\times SU(2)$ Yang-Mills multiplets. The consistency of the reduction is nontrivial and surprising, since the analogous reduction of type IIB on $T^{1,1}$ was shown to be not consistent, despite the fact that AdS$_5\times T^{1,1}$ is its vacuum \cite{Hoxha:2000jf}. It can be argued that without our unusual duality between the heterotic and bosonic strings, we would not have considered the possibility of making consistent Pauli-type reduction of the heterotic theory on this internal space that involves $T^{1,1}$, even if we had constructed the new vacuum by other means.

It is worth noting that the bosonic field contents also make up precisely those of ${\cal N}=4$, $D=4$ gauged supergravity with $SO(4)$ gauge group \cite{zwiebach,Gates:1983ha}, which can be obtained from the nontrivial $S^7$ reduction of 11-dimensional supergravity \cite{Cvetic:1999au}. In addition, there exists an inequivalent Freedman-Schwarz model \cite{Freedman:1978ra} with $SU(2)\times SU(2)$ gauge fields whose origin is ${\cal N}=1$, $D=10$ pure supergravity on $S^3\times S^3$ \cite{Cowdall:1998bu, Chamseddine:1998km}.

There are some similarities of our supergravity and the Freedman-Schwarz model, as they both have a complex scalar and $SU(2)\times SU(2)$ gauge symmetry, and they both have origins in 10 dimensions. However, the Freedman-Schwarz model is gauged with a running dilaton scalar potential and hence the Minkowski spacetime is not its vacuum. Its 10-dimensional origin is the minimum supergravity, with no matter field involved. The 10-dimensional origin of our theory, on the other hand, requires an additional Maxwell matter field. Furthermore, our theory has less (minimum) supersymmetry and it has Minkowski vacuum. These properties make it more interesting in studying its phenomenology.

\section*{Acknowledgement}

We are grateful to Chris Pope for useful comments. L.M.~is supported in part by National Natural Science Foundation of China (NSFC) grant No.~12247103, Postdoctoral Fellowship Program of CPSF Grant No.~GZC20241211 and the China Postdoctoral Science Foundation under Grant No.~2024M762338. H.L.~is supported in part by the National Natural Science Foundation of China (NSFC) grants No.~12375052 and No.~11935009.

\appendix
\subsection*{Appendices}
\section{Useful formulae}
\label{app:formulae}

In this appendix, we present useful formulae associated with the 2-sphere, 3-sphere and their relations. (See also \cite{Gibbons:2003gp}.) Consider a unit round $S^2$ parameterized by $\mu^\alpha \mu^\alpha =1$, with
\be
\mu^1 = \sin\theta \sin\psi\,,\qquad
\mu^2 = \sin\theta \cos\psi\,,\qquad
\mu^3 = \cos\theta\,.
\ee
Its metric is given by
\be
ds_2^2 = d\theta^2 + \sin^2\theta d\psi^2= d\mu^\alpha d\mu^\alpha = e^a e^a\,,
\ee
The three Killing vectors are given by
\be
K_\alpha^m =\epsilon^{mn} \partial_n \mu_\alpha\,.
\ee
Thus we have
\bea
K_1 = \cos\psi\, \fft{\partial}{\partial\theta} - \sin\psi \cot\theta\, \fft{\partial}{\partial\psi}\,,\quad
K_2 = - \sin\psi\, \fft{\partial}{\partial\theta} - \cos\psi \cot\theta\,
\fft{\partial}{\partial\psi}\,,\quad
K_3 = \fft{\partial}{\partial\psi}\,.
\eea
We have various useful identities
\be
K_\alpha^m K_\alpha^n = g^{mn}\,,\qquad g_{mn} K_\alpha^m K_\beta^n = \delta_{\alpha\beta} - \mu_\alpha \mu_\beta\,,\qquad
K_\alpha^a e^a = -\epsilon_{\alpha\beta\gamma} \mu^\beta d\mu^\gamma\,.
\ee
\be
K_\alpha = -\epsilon_{\alpha\beta\gamma} \mu^\beta \wedge d\mu^\gamma\,,\qquad
dK_\alpha = 2\mu^\alpha \Omega_\2 = - \epsilon_{\alpha\beta\gamma} d\mu^\beta
\wedge d\mu^\gamma\,.
\ee
Note that $S^3$ can be viewed as a $U(1)$ bundle of $S^2$.  We can thus rewrite
$h^\alpha$, defined by \eqref{hdef}, by
\bea
h^\alpha&=&-\epsilon^{\alpha\beta\gamma}\mu_\beta D\mu_\gamma+\mu^\alpha\sigma\,,\qquad D\mu^\alpha=d\mu^\alpha+g\epsilon^{\alpha\beta\gamma}A_{\1\beta}\mu_\gamma\,,\cr
\sigma &=&  d\varphi+\mathcal{A}_{\1}\,,\qquad \mathcal{A}_{\1}=\cos\theta d\psi-g\mu_\alpha A_{\1}^\alpha\,.
\eea
The metric $h_\alpha h^\alpha$ can be expressed as Kaluza-Klein form
\bea
h_\alpha h^\alpha&=&D\mu^\alpha D\mu_\alpha+\sigma^2,\cr
D\mu^\alpha D\mu_\alpha&=&(d\theta-gA^1_{\1}\cos\psi+gA^2_{\1}\sin\psi)^2\cr
&&+\sin^2\theta(d\psi+gA^1_{\1}\cot\theta\sin\psi+gA^2_{\1}
\cot\theta\cos\psi-gA^3_{\1})^2=\varepsilon^a \varepsilon^a.
\eea
Here
\be
\varepsilon^a = e^a - g K^a_\alpha A^\alpha\,.\label{varep}
\ee
An important identity for form fields is that
\be
\ft12 \epsilon_{\alpha\beta \gamma} \mu^\alpha D\mu^\beta\wedge D\mu^\gamma = -\ft12 \epsilon_{ab}\, \varepsilon^a\wedge\varepsilon^b\,.
\ee
These formulae are also valid for the $S^3$ with tilded coordinates.

\section{Generalize to $D$ dimensions}
\label{app:gend}

In the main text, we discussed a duality between the bosonic sector of the heterotic gravity and anomalous string theory in ten dimensions. This discussion can be generalized to arbitrary dimensions.

We first consider the heterotic-inspired theory in general $D$ dimensions, consisting of the same bosonic field contents as the heterotic theory. The Lagrangian is given by
\be
\hat{\mathcal{L}}_{D}= \hat{R}\, {\hat *\oneone}-\ft12 {\hat * d\hat \phi}\wedge d\hat \phi -\ft12 e^{a_D\hat\phi} {\hat *\hat H_\3} \wedge \hat H_\3 -\ft{1}{2 }e^{\frac{1}{2} a_D\hat{\phi}}{\hat *\hat F_\2}\wedge \hat F_\2\,,\label{heteroticD}
\ee
where $\hat{F}_{\2}=d\hat{A}_{\1}$, $\hat{H}_{\3}=d\hat{B}_{\2} +\frac{1}{2}\hat{F}_{\2}\wedge \hat{A}_{\1}$ and
\be
a_D=-\sqrt{\fft{8}{D-2}}\,.
\ee
We consider the warped $\mathbb R$ reduction ansatz on the internal coordinate $z$:
\bea
d\hat{s}_{D}^2&=&e^{2mz}\Big(
e^{-\sqrt{\fft{2}{D-3}}\fft{\bar\phi}{D-2}}\,d\bar{s}_{D-1}^2
+e^{\fft{\sqrt{2(D-3)}}{D-2}\bar\phi}\,dz^2
\Big),\cr
\hat{\phi}&=&-\sqrt{\fft{D-3}{D-2}}\,\bar{\phi}-4mz\,,\qquad \hat{B}_{\2}=\bar{B}_{\2}\,,\qquad \hat{A}_{\1}=\bar{A}_{\1}\,.\label{genwarpedr}
\eea
Here, $m$ is an arbitrary constant. It can be verified that the reduction is consistent and we obtain the $(D-1)$-dimensional Lagrangian
\be
\bar{\mathcal{L}}_{D-1}=
\big(\bar{R} -64m^2e^{-\frac{1}{2}a_{D-1}\bar{\phi}}\big){\bar *\oneone}
-\ft12 {\bar *d\bar \phi}\wedge d\bar\phi -\ft12 e^{a_{D-1}\bar{\phi}}
{\bar * \bar H}_\3\wedge \bar H_\3
-\ft12 e^{\frac{1}{2}a_{D-1}\bar{\phi}} {\bar *\bar F}_\2\wedge \bar F_\2\,,\label{dm1lag}
\ee
where $\bar{H}_{\3}=d\bar{B}_{\2}+\frac{1}{2}\bar{F}_{\2}\wedge \bar{A}_{\1}$ and $\bar{F}_{\2}=d\bar{A}_{\1}$. The Lagrangian of the anomalous bosonic string theory in general $D$ dimensions takes the form
\bea
\check{\mathcal{L}}_{D}=
(\check{R}-8g^2 e^{-\frac{1}{2} a_{D}\check{\phi}})\,{\check * \oneone} -\ft12 {\check * d\check \phi}\wedge d\check\phi-\frac{1}{2}
e^{a_{D} \check{\phi}}{\check*\check{H}}_\3\wedge \check H_\3 \,,\qquad
\check{H}_{\3}=d\check{B}_{\2}\label{gendstr}
\eea
We now consider $S^1$ reduction of this theory, but keeping the winding and Kaluza-Klein modes equal. In other words, we have
\bea
d\check{s}_D^2&=&e^{-\sqrt{\frac{2}{D-3}}\frac{\bar \phi}{D-2}}\,d \bar s_{D-1}^2+e^{\frac{\sqrt{2(D-3)}}{D-2}\bar \phi}\, (dz-\frac{1}{\sqrt{2}}\bar A_\1)^2\,,\cr
\check{B}_\2&=&\bar B_\2+\frac{1}{\sqrt{2}}\bar A_\1\wedge dz\,,\qquad
\check{\phi}=-\sqrt{\frac{D-3}{D-2}}\,\bar\phi\,.\label{s1red}
\eea
We find that the $D$-dimensional Lagrangian \eqref{gendstr} can be consistently reduced on this ansatz to the same $(D-1)$-dimensional Lagrangian \eqref{dm1lag}, provided that the parameters $m$ and $g$ satisfy the (dimensional-independent) relation \eqref{mgrelation}.

We can perform the consistent $S^3\times S^3$ DeWitt type reduction of the bosonic string \eqref{gendstr}, with the metric reduction ansatz
\be
d\check{s}_D^2=e^{\sqrt{\frac{2}{D-8}}\frac{6}{D-2}\phi}\,ds_{D-6}^2
+\frac{1}{4g^2}e^{-\frac{\sqrt{2(D-8)}}{D-2}\phi}\big(h^\alpha h_\alpha+\tilde{h}^{\tilde{\alpha}} \tilde{h}_{\tilde{\alpha}}\big).
\ee
The reduction of the scalar field is $\check{\phi}=\sqrt{\frac{D-8}{D-2}}\phi$.
The form field reductions are independent of the dimensions, taking the same forms as those from $D=10$ to $D=9$, as discussed in the main text.

The duality mapping above allows us to obtain the ansatz of the heterotic-inspired theory in general dimensions on the warped $\mathbb R\times T^{1,1}$ internal space. Again the reduction on the form fields are the same as before, independent of the dimensions. The reductions on the metric and the scalar are
\bea
d\hat s_{D}^2 &=& e^{2mz}\Bigg(e^{\sqrt{\frac{2}{D-8}}\frac{6}{D-2}\phi}\, ds_{D-6}^2 +e^{-\frac{\sqrt{2(D-8)}}{D-2}\phi}\Big[ dz^2\nn\\
&&+\frac{1}{8g^2}\left(\Sigma-g\mu_\alpha A_{\1}^\alpha -g\tilde{\mu}_{\tilde{\alpha}} \widetilde{A}_{\1}^{\tilde{\alpha}}\right)^2+\frac{1}{4g^2}\left(D\mu^\alpha D\mu_\alpha+D\tilde{\mu}^{\tilde{\alpha}} D\tilde{\mu}_{\tilde{\alpha}}\right)\Big]\Bigg)\,.\nn\\
\hat \phi &=& \sqrt{\frac{D-8}{D-2}} \phi - 4mz\,,
\eea
In both cases, we obtain the same $(D-6)$-dimensional theory
\bea
\mathcal{L}_{D-6}&=&R_{D-6}\, {*\oneone}-\ft{1}{2} {* d\phi}\wedge d\phi-\ft{1}{2}e^{-a_{D-6}\phi} {*H}_{\3}\wedge H_{\3}\cr
&&-\ft{1}{4}e^{-\frac{1}{2}a_{D-6}\phi}\big( {* F}_{\2}^\alpha\wedge F_{\2\alpha} + {*\widetilde{F}}_{\2}^{\tilde{\alpha}}\wedge \widetilde{F}_{\2\tilde{\alpha}}\big).\label{dm6theory}
\eea
Taking $D=10$ yields the four-dimensional theory discussed in the main text.

\section{Salam-Sezgin model on $S^2$ from bosonic string on $S^3$}
\label{app:3=2+1}

In this section, we derive the $S^2$ Pauli reduction ansatz \cite{Gibbons:2003gp} of the bosonic sector of the Salam-Sezgin model. We make use of the observation \cite{Ma:2023rdq} that the bosonic sector of the Salam-Sezgin model can be also obtained from the anomalous bosonic string on $S^1$. The discussion can in fact be made in general dimensions, since the
$(D-1)$-dimensional Lagrangian \eqref{dm1lag} becomes precisely the bosonic sector the Salam-Sezgin model when we take $D=7$. Thus we can start with the simpler DeWitt type of $S^3$ reduction of the anomalous bosonic string \eqref{gendstr}. We perform $S^1$ reduction and obtain the $S^2$ reduction ansatz of \eqref{dm1lag}.

\subsection{$S^3$ reduction of the anomalous bosonic string}

The reduction ansatz of this $SU(2)$ truncation of the original $SO(4)$ Kaluza-Klein reduction is \cite{Cvetic:2000dm}
\bea
d\check{s}_D^2&=&X^{\frac{6}{D-2}}ds_{D-3}^2+\frac{1}{4g^2}
X^{-\frac{2(D-5)}{D-2}}h^\alpha h_\alpha\,,\cr
\check{B}_{\2}&=&B_{\2}+\frac{1}{4g}(\cos\psi A_{\1}^1-\sin\psi A_{\1}^2)\wedge d\theta+\frac{1}{4g}A_{\1}^3\wedge d\psi\cr
&&-\frac{1}{4g^2}(\cos\theta d\psi-g\mu_\alpha A_{\1}^\alpha)\wedge d\varphi\,,\cr
e^{\sqrt{\frac{D-2}{2}}\check{\phi}}&=&X^{D-5}\,.\label{S3 reduction SU2}
\eea
We can also use the field strength reduction to express the ansatz
\bea
\check{H}_{\3}&=&H_{\3}-\frac{1}{4g^2}\Omega_{\3}+
\frac{1}{4g} F_{\2}^\alpha\wedge h_\alpha\,,\cr
H_{\3}
&=&dB_{\2}+\frac{1}{4} F_{\2}^\alpha\wedge A_{\1\alpha}-\frac{g}{24}\epsilon_{\alpha\beta\gamma}A_{\1}^\alpha\wedge A_{\1}^\beta\wedge A_{\1}^\gamma\,,\cr
F_{\2}^\alpha&=&dA_{\1}^\alpha+\frac{g}{2}\epsilon^{\alpha\beta\gamma}
A_{\1\beta}\wedge A_{\1\gamma}\,.
\eea
The $S^3$ volume form is $\omega_{\3}=\frac{1}{8}\Omega_{\3}$, where
\be
\Omega_{\3}=\frac{1}{6}\epsilon_{\alpha\beta\gamma}h^\alpha\wedge h^\beta\wedge h^\gamma =h^1\wedge h^2\wedge h^3\,.
\ee
This DeWitt-type of $S^3$ reduction is consistent and we obtain $(D-3)$-dimensional Einstein gravity, coupled to an $SU(2)$ Yang-Mills field, a scalar and a 3-form field strength. The lower-dimensional Lagrangian is
\be
\mathcal{L}_{D-3}=R_{D-3}\,{*\oneone}-(D-5)X^{-2}{* dX}\wedge dX-\ft{1}{2}X^{-4} {* H}_{\3}\wedge H_{\3}-\ft{1}{4}X^{-2}
{* F}_{\2}^\alpha\wedge F_{\2\alpha}\,. \label{dm3theory}
\ee

\subsection{$S^2$ reduction of Salam-Sezgin model}

We now impose the relation between Killing coordinate $\varphi$ in $S^3$ reduction \eqref{S3 reduction SU2} and $z$ in circle reduction \eqref{s1red}
\be
\varphi= 2gz\,.
\ee
We can compare the $S^3$ reduction \eqref{S3 reduction SU2} and the circle reduction \eqref{s1red}, making use of the formulae in appendix \ref{app:formulae}, we obtain the $S^2$ reduction ansatz from $(D-1)$ to $(D-3)$ dimensions:
\bea
d\bar{s}_{D-1}^2&=&
X^{\frac{4}{D-3}}ds_{D-3}^2+\frac{1}{4g^2}X^{-\frac{2(D-5)}{D-3}}D\mu^\alpha D\mu_\alpha,\cr
D\mu^\alpha D\mu_\alpha&=&(d\theta-gA^1_{\1}\cos\psi+gA^2_{\1}\sin\psi)^2\cr
&&+\sin^2\theta(d\psi+gA^1_{\1}\cot\theta\sin\psi+gA^2_{\1}
\cot\theta\cos\psi-gA^3_{\1})^2\,,\cr
\bar{A}_{\1}&=&-\frac{1}{\sqrt{2}g}\cos\theta d\psi+\frac{1}{\sqrt{2}}\mu_\alpha A_{\1}^\alpha\,,\cr
\bar{B}_{\2}&=&B_{\2}+\frac{1}{4g}(\cos\psi A_{\1}^1-\sin\psi A_{\1}^2)\wedge d\theta+\frac{1}{4g}A_{\1}^3\wedge d\psi\,,\cr
e^{\sqrt{\frac{D-3}{2}}\frac{\bar{\phi}}{D-5}}&=&X.\label{s2red}
\eea
Under this reduction ansatz, the $(D-1)$-dimensional Lagrangian \eqref{dm1lag} consistently reduces to the $(D-3)$-dimensional theory \eqref{dm3theory}. When we take $D=7$, the theory \eqref{dm1lag} becomes the bosonic sector of the Salam-Sezgin model, and the reduction ansatz \eqref{s2red} become the one given in \cite{Gibbons:2003gp}.

Finally, there is an convention alert. We adopted the same convention of \cite{Cvetic:2000dm} for the $S^3$ reduction of the bosonic string, in which case, the conformal anomaly term in \eqref{gendstr} should be $4g^2$ instead of the $8g^2$ adopted in this paper and also in \cite{Gibbons:2003gp}. With this understood, then sending $A^\alpha_\1 \rightarrow \sqrt2 A^\alpha_\1$ and $g\rightarrow \sqrt2 g$, the theory \eqref{dm3theory} in $D=4$ becomes precisely the one in \cite{Gibbons:2003gp}, with the same convention of the kinetic term and gauge coupling.

\end{document}